\documentstyle[epsfig,psfig]{texas}

\newcommand{\JVAS}{{\sf JVAS}}
\newcommand{\CLASS}{{\sf CLASS}}
\newcommand{\CERES}{{\sf CERES}}
\def\sign{\mathop{\rm sign}\nolimits}
\begin{document}

\title{Measuring Cosmological Parameters with the \JVAS\ and \CLASS\ Gravitational 
Lens Surveys}

\author{
P.~Helbig$^{1}$,
R.~D.~Blandford$^{2}$,
I.~W.~A.~Browne$^{1}$,
A.~G.~de~Bruyn$^{3,4}$,
C.~D.~Fassnacht$^{2,5}$,
N.~Jackson$^{1}$,
L.~V.~E.~Koopmans$^{4}$,
J.~F.~Macias-Perez$^{1}$,
D.~R.~Marlow$^{1,6}$,
S.~T.~Myers$^{6}$,
R.~Quast$^{7}$,
D.~Rusin$^{6}$,
P.~N.~Wilkinson$^{1}$,
E.~Xanthopoulos$^{1}$
}

\address{(1) 
University of Manchester, Nuffield Radio Astronomy Laboratories\\ 
Jodrell Bank, Macclesfield, SK11 9DL, UK
}
\address{(2) 
California Institute of Technology\\
Pasadena, CA 91125, USA
}
\address{(3) 
Netherlands Foundation for Research in Astronomy\\
P.~O.~Box 2, NL-7990 AA Dwingeloo, The Netherlands
}
\address{(4)
University of Groningen, Kapteyn Astronomical Institute\\
Postbus 800, NL-9700 AV Groningen,
The Netherlands
}
\address{(5)
National Radio Astronomy Observatory\\
P.O. Box 0, Socorro, NM 87801, USA
}
\address{(6)
University of Pennsylvania, Dept.~of Physics and Astronomy\\
209 S.~$33^{\rm rd}$ Street, Philadelphia, PA, 19104-6394, USA
}
\address{(7) University of Hamburg, Hamburg Observatory\\
Gojenbergsweg 112, 21029 Hamburg, Germany
}

\begin{abstract}
The \JVAS\ (Jodrell Bank-VLA Astrometric Survey) and \CLASS\ 
(Cosmic Lens All-Sky Survey) are well-defined surveys containing about ten
thousand flat-spectrum radio sources.  For many reasons, flat-spectrum
radio sources are particularly well-suited as a population from which 
one can obtain
unbiased samples of gravitational lenses.  These are by far the
largest gravitational (macro)lens surveys, and particular attention was
paid to constructing a cleanly-defined sample for the survey itself
and for the underlying luminosity function.  Here we present the
constraints on cosmological parameters, particularly the cosmological
constant, derived from \JVAS\ and combine them with constraints from optical
gravitational lens surveys, `direct' measurements of $\Omega_{0}$,
$H_{0}$ and the age of the universe, and constraints derived from CMB 
anisotropies, before putting this final result into the context of the 
latest results from other, independent cosmological tests.
\end{abstract}

%%%%%%%%%%%%%%%%%%%%%%%%%%%%%%%%%%%%%%%%%%%%%%%%%%%%%%%%%%%%%%%%%%%%%%%%
%%%%%%%%%%%%%%%%%%%%%%%%%%%%%%%%%%%%%%%%%%%%%%%%%%%%%%%%%%%%%%%%%%%%%%%%
%%%%%%%%%%%%%%%%%%%%%%%%%%%%%%%%%%%%%%%%%%%%%%%%%%%%%%%%%%%%%%%%%%%%%%%%
\section{Cosmological constraints from \JVAS\dots}

The Jodrell Bank-VLA Astrometric Survey (\JVAS) is a survey for
flat-spectrum radio sources with a flux density greater than 200\,mJy at
5\,GHz. Flat-spectrum radio sources are likely to be compact, thus
making it easy to recognise the lensing morphology.  In addition, they
are likely to be variable, making it possible to determine $H_{0}$ by
measuring the time delay between the lensed images.  (See
\cite{ABiggsBHKWP99a} for the description of a time delay measurement
in a \JVAS\ gravitational lens system.)  \JVAS\ is also a survey for
MERLIN phase-reference sources and as such is described in
\cite{APatnaikBWW92a}, \cite{IBRownePWW98a} and \cite{PWilkinsonBPWS98a}.
\JVAS\ as a gravitational lens survey, the lens candidate selection, followup
process, confirmation criteria and a discussion of the \JVAS\
gravitational lenses is described in detail in \cite{LKingBMPW99a}
(see also \cite{LKingIBrowne96a}).

In order to have a parent sample which is as large as possible and as
cleanly defined as practical, our `\JVAS\ gravitational lens survey
sample' is slightly different than the `\JVAS\ phase-reference calibrator
sample'.  For the former, the source must be a point source and must
have a good starting position (so that the observation was correctly
pointed) while its precise spectral index is not important.  For the
latter, only the spectral index is important, as the source can be
slightly resolved or the observation can be less than perfectly pointed.
Thus, the \JVAS\ astrometric sample
\cite{APatnaikBWW92a,IBRownePWW98a,PWilkinsonBPWS98a} contains 2144
sources.  To these must be added 103 sources which were too resolved to
be used as phase calibrators and 61 sources which had bad starting
positions (thus the observations were too badly pointed to be useful for
the astrometric sample), bringing the total to 2308.  This formed our
gravitational lens sample, since these additional sources were also
searched for
gravitational lenses \cite{LKingBMPW99a} (none were found meeting the
\JVAS\ selection criteria: multiple flat-spectrum point-source 
components with a separation between 300 mas and 6 arcsec with a flux 
ratio of $\le$20).

We have used the gravitational lens systems in Table~\ref{ta:lenses} in 
this analysis.  
%%%%%%%%%%%%%%%%%%%%%%%%%%%%%%%%%%%%%%%%%%%%%%%%%%%%%%%%%%%%%%%%%%%%%%%
%%%%%%%%%%%%%%%%%%%%%%%%%%%%%%%%%%%%%%%%%%%%%%%%%%%%%%%%%%%%%%%%%%%%%%%
%\begin{table}*
\begin{table}[h]
\begin{center}
\begin{tabular*}{\textwidth}{@{\extracolsep{\fill}}llllll}
\hline
\hline
Name & 
\# images & 
$\Delta\theta ['']$ & 
$z_{\mathrm{l}}$ & 
$z_{\mathrm{s}}$ & 
lens galaxy \\
\hline
%_________________________________________________________________
B0218+357     &                % NAME
2 + ring      &                % # IMAGES
0.334         &                % IMAGE SEPARATION
0.6847        &                % LENS REDSHIFT
0.96          &                % SOURCE REDSHIFT
spiral        \\               % LENS GALAXY TYPE
%_________________________________________________________________
MG0414+054    &                % NAME
4             &                % # IMAGES
2.09          &                % IMAGE SEPARATION
0.9584        &                % LENS REDSHIFT
2.639         &                % SOURCE REDSHIFT
elliptical    \\               % LENS GALAXY TYPE
%_________________________________________________________________
B1030+074     &                % NAME
2             &                % # IMAGES
1.56          &                % IMAGE SEPARATION
0.599         &                % LENS REDSHIFT
1.535         &                % SOURCE REDSHIFT
spiral        \\               % LENS GALAXY TYPE
%_________________________________________________________________
B1422+231     &                % NAME
4             &                % # IMAGES
1.28          &                % IMAGE SEPARATION
0.337         &                % LENS REDSHIFT
3.62          &                % SOURCE REDSHIFT
\textbf{?}    \\               % LENS GALAXY TYPE
\hline
\end{tabular*}
\caption[]{
%\Large\bf\boldmath
\JVAS\ lenses used in this analysis.  Of the information in the 
table, for this analysis we use only the source redshift $z_{\mathrm{s}}$,
and the image separation $\Delta\theta$.}
\label{ta:lenses}
%\end{table*}
\end{center}
\end{table}
%%%%%%%%%%%%%%%%%%%%%%%%%%%%%%%%%%%%%%%%%%%%%%%%%%%%%%%%%%%%%%%%%%%%%%%
The \JVAS\ lens B1938+666 \cite{LKingJBBBdBFKMNW98a} was not included 
because it 
is not formally a part of the sample, having a too steep spectral index and 
having been recognised on the basis of a lensed extended 
source as opposed to lensed compact components.  Also, the \JVAS\ lens 
B2114+022 \cite{PAugustoBWJFM99a} was not 
included because it is not a single-galaxy lens system.

\begin{figure}[t]
\centering
\epsfig{figure=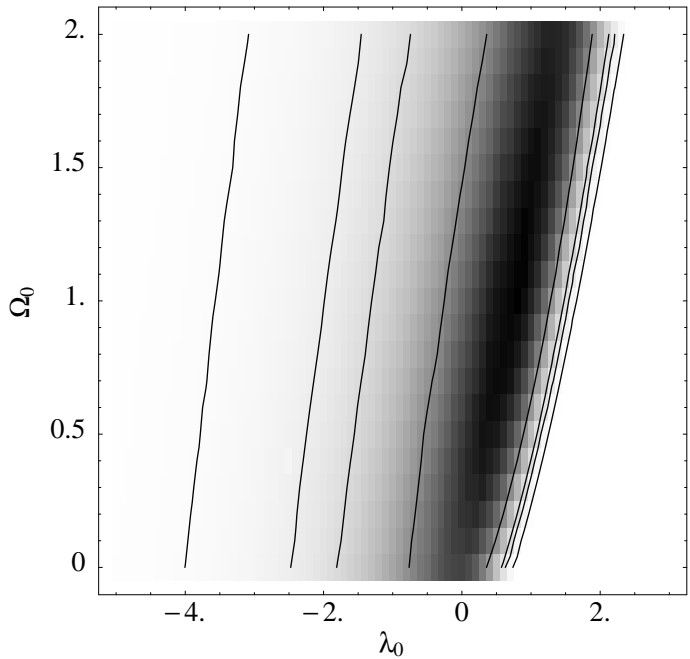, width=9cm}
\caption[]{
%\Large\bf\boldmath
The likelihood function $p(\lambda_0,\Omega_0)$ based on
the \JVAS\ lens sample.  The pixel grey level is directly proportional to
the likelihood: darker pixels reflect higher likelihoods.  The pixel
size reflects the resolution of our numerical computations.  The
contours mark the boundaries of the minimum $0.68$, $0.90$, $0.95$ and
$0.99$ confidence regions for the parameters $\lambda_0$ and $\Omega_0$.}
\label{fi:jvas}
\end{figure}

For this analysis, due to the paucity of the observational data, we have
made rather stark assumptions: the redshift distribution of the sample
is assumed to be identical to that of the CJF (Caltech-Jodrell Bank 
Flat-spectrum) sample
\cite{GTaylorVRPHW96a},
independent of flux density, and the number-magnitude relation is 
assumed to be
identical to that of \CLASS\, the
Cosmic Lens All-Sky Survey, \cite{SMyersetal99a}, independent of
redshift.  Otherwise, we have calculated the likelihood as a function of
$\lambda_{0}$ and $\Omega_{0}$ as described in
\cite{CKochanek96a}.  That is, we use a non-singular isothermal sphere 
as a lens model, model the lens galaxy population with a Schechter 
function and use the Faber-Jackson relation to convert between 
luminosity and velocity dispersion, considering only elliptical galaxies.
The results are presented in Fig.~\ref{fi:jvas}.
%%%%%%%%%%%%%%%%%%%%%%%%%%%%%%%%%%%%%%%%%%%%%%%%%%%%%%%%%%%%%%%%%%%%%%%%
%%%%%%%%%%%%%%%%%%%%%%%%%%%%%%%%%%%%%%%%%%%%%%%%%%%%%%%%%%%%%%%%%%%%%%%%
%%%%%%%%%%%%%%%%%%%%%%%%%%%%%%%%%%%%%%%%%%%%%%%%%%%%%%%%%%%%%%%%%%%%%%%%
At 95\% confidence, our lower and upper limits on
$\lambda_{0}-\Omega_{0}$, using the \JVAS\ lensing statistics 
information
alone, are
respectively $-2.69$ and $0.68$.  For a flat universe, these correspond
to lower and upper limits on $\lambda_{0}$ of respectively $-0.85$ and
$0.84$.  (Reducing the constraints in the $\lambda_{0}$-$\Omega_{0}$ 
plane to $\lambda_{0}-\Omega_{0}$ is, of course, just an approximation, 
but a reasonably good one when considering upper limits on $\lambda_{0}$
for small $\Omega_{0}$ values.  These numbers were derived from the 
corresponding confidence limits on $\lambda_{0}$ for fixed $\Omega_{0}$
and are thus of course different than the intersection of lines of 
constant $\lambda_{0}-\Omega_{0}$ with the corresponding contour in 
Fig.~\ref{fi:jvas}.)

%%%%%%%%%%%%%%%%%%%%%%%%%%%%%%%%%%%%%%%%%%%%%%%%%%%%%%%%%%%%%%%%%%%%%%%%
%%%%%%%%%%%%%%%%%%%%%%%%%%%%%%%%%%%%%%%%%%%%%%%%%%%%%%%%%%%%%%%%%%%%%%%%
%%%%%%%%%%%%%%%%%%%%%%%%%%%%%%%%%%%%%%%%%%%%%%%%%%%%%%%%%%%%%%%%%%%%%%%%
\section{\dots and optical gravitational lens 
surveys\dots}

\begin{figure}[t]
\centering
\epsfig{figure=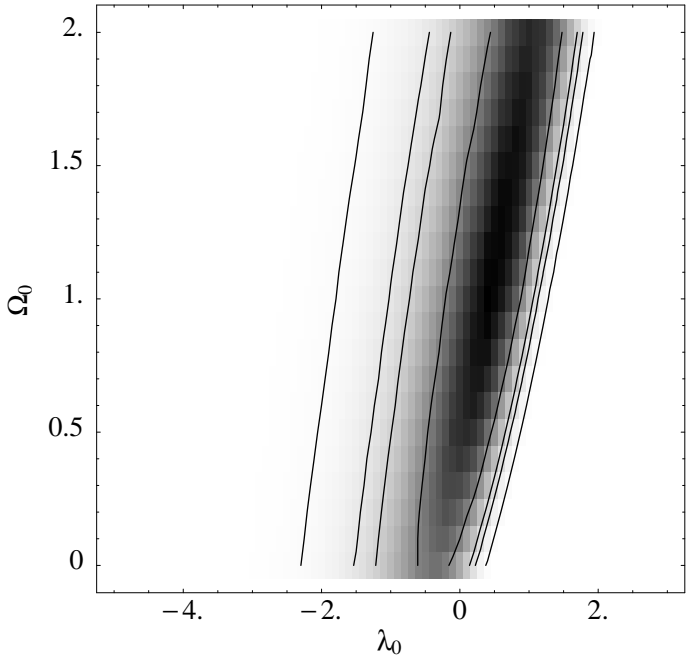, width=9cm}
\caption[]{
%\Large\bf\boldmath
The same as Fig.~\ref{fi:jvas} but combining 
\JVAS\ with optical 
gravitational lens surveys from the literature.}
\label{fi:joint}
\end{figure}

One can improve these constraints by adding those from optical 
gravitational lens surveys, though one should keep in mind that the 
systematic errors---for example, lens systems which are missed due to
extinction in the lens galaxy or to the fact that the typical seeing
is not much better than the typical image separation---are probably less 
well understood than is the case in 
the radio (though the statistical properties of the unlensed parent 
population are better understood).  Not only does one have more objects 
and thus better statistics, but a different redshift range is sampled as 
well.  Essentially repeating the analysis in \cite{CKochanek96a} (but 
with $\lambda_{0}$ and $\Omega_{0}$ as free parameters, of course) and 
combining the resulting constraints with those from Fig.~\ref{fi:jvas}, 
one obtains the better constraints shown in Fig.~\ref{fi:joint}. 
%%%%%%%%%%%%%%%%%%%%%%%%%%%%%%%%%%%%%%%%%%%%%%%%%%%%%%%%%%%%%%%%%%%%%%%%
%%%%%%%%%%%%%%%%%%%%%%%%%%%%%%%%%%%%%%%%%%%%%%%%%%%%%%%%%%%%%%%%%%%%%%%%
%%%%%%%%%%%%%%%%%%%%%%%%%%%%%%%%%%%%%%%%%%%%%%%%%%%%%%%%%%%%%%%%%%%%%%%%
Using the combination of \JVAS\ lensing statistics and lensing statistics
from the literature as in \cite{CKochanek96a}, the
corresponding $\lambda_{0}-\Omega_{0}$ values are
$-1.78$ and $0.27$.  For a
flat universe, these correspond
to lower and upper limits on $\lambda_{0}$ of respectively $-0.39$
and $0.64$.

%%%%%%%%%%%%%%%%%%%%%%%%%%%%%%%%%%%%%%%%%%%%%%%%%%%%%%%%%%%%%%%%%%%%%%%%
%%%%%%%%%%%%%%%%%%%%%%%%%%%%%%%%%%%%%%%%%%%%%%%%%%%%%%%%%%%%%%%%%%%%%%%%
%%%%%%%%%%%%%%%%%%%%%%%%%%%%%%%%%%%%%%%%%%%%%%%%%%%%%%%%%%%%%%%%%%%%%%%%
\section{\dots and `reasonably well-accepted 
wisdom'\dots}

Gravitational lensings statistics alone cannot usefully constrain 
$\Omega_{0}$.  Thus, it seems sensible to combine the constraints shown 
in Fig.~\ref{fi:joint} with measurements of $\Omega_{0}$.  Fortunately,
there seems to be a consensus developing that $\Omega_{0} \approx 0.3$
e.g.~\cite{RCarlbergetal97a,RCarlberg98a,RCarlbergetal98b,
NBahcall97a,NBahcallFC97a,XFanBC97a,MBartelmannHCJP98a,
SPerlmutteretal98a,
ARiessetal98a,BSchmidtetal98a,AKim98a,CLineweaver98a,EGuerraDW98a,
RDalyGW98a}.  Conservatively, 
these results can be summarised as
\begin{equation}
  p(\lambda_0,\Omega_0) =
  L(\Omega_0|0.4,0.2).
  \label{eq:p2}
\end{equation}
where the two arguments of $L$ represent the mean and standard deviation 
of a lognormal distribution.

In a similar vein, lensing statistics determines a lower limit on 
$\lambda_{0}$ much less strongly than an upper limit, so it seems sensible to
include some prior information which can give a lower limit on 
$\lambda_{0}$.  To be conservative, we take relatively undisputed 
estimates for the age of the universe and the Hubble constant, their 
product setting a (slightly $\Omega_{0}$-dependent) lower limit on 
$\lambda_{0}$.
The best estimate of the absolute age of the oldest galactic globular
clusters currently is $t_{\mathrm{gc}}=11.5\pm1.3\,\mbox{Gyr}$
\cite{BChaboyerDKK98a}. We choose to formulate this prior information
in the form of a lognormal distribution that meets these statistics
\begin{equation}
  p(t_{\mathrm{gc}}) =
  L(t_{\mathrm{gc}}|11.5\,\mbox{Gyr}, 1.3\,\mbox{Gyr}).
  \label{eq:ptgc}
\end{equation}
Similarly, we roughly estimate
$H_0=65\pm10\,\mbox{km}\,\mbox{s}^{-1}\,\mbox{Mpc}^{-1}$ and choose to
formulate this prior information in form of a normal distribution
\begin{equation}
  p(H_0) =
  N(H_0|65\,\mbox{km}\,\mbox{s}^{-1}\,\mbox{Mpc}^{-1},
    10\,\mbox{km}\,\mbox{s}^{-1}\,\mbox{Mpc}^{-1}),
  \label{eq:ph0}
\end{equation}
where, again, the notation for $L$ (and $N$) is such that the two arguments
correspond to
the mean and standard deviation.

\begin{figure}[t]
\centering
\epsfig{figure=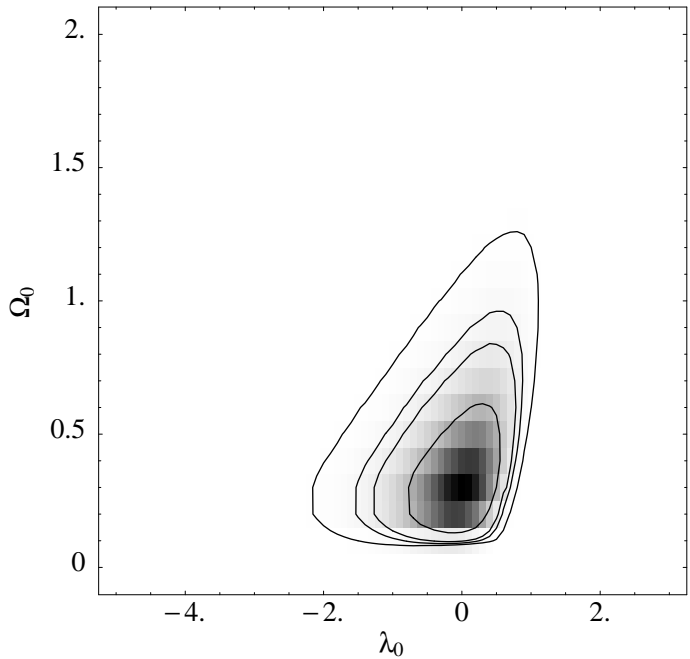, width=9cm}
\caption[]{
%\Large\bf\boldmath
The same as Fig.~\ref{fi:joint} but 
combining \JVAS\ and optical 
gravitational lens surveys from the literature with prior information on 
the value of $\Omega_{0}$, $H_{0}$ and the age of the universe.  This 
figure thus represents the combination of constraints from lensing 
statistics and from relatively undisputed knowledge about values of the 
cosmological parameters.}
\label{fi:posterior}
\end{figure}

Fig.~\ref{fi:posterior} shows how inclusion of this prior information, 
representing a conservative estimate of what we know about the values of 
the cosmological parameters, tightens the constraints on $\lambda_{0}$ 
and $\Omega_{0}$ as compared to the constraints from lens statistics 
alone (Figs.~\ref{fi:joint} and \ref{fi:jvas}).
%%%%%%%%%%%%%%%%%%%%%%%%%%%%%%%%%%%%%%%%%%%%%%%%%%%%%%%%%%%%%%%%%%%%%%%%
%%%%%%%%%%%%%%%%%%%%%%%%%%%%%%%%%%%%%%%%%%%%%%%%%%%%%%%%%%%%%%%%%%%%%%%%
%%%%%%%%%%%%%%%%%%%%%%%%%%%%%%%%%%%%%%%%%%%%%%%%%%%%%%%%%%%%%%%%%%%%%%%%

%%%%%%%%%%%%%%%%%%%%%%%%%%%%%%%%%%%%%%%%%%%%%%%%%%%%%%%%%%%%%%%%%%%%%%%%
%%%%%%%%%%%%%%%%%%%%%%%%%%%%%%%%%%%%%%%%%%%%%%%%%%%%%%%%%%%%%%%%%%%%%%%%
%%%%%%%%%%%%%%%%%%%%%%%%%%%%%%%%%%%%%%%%%%%%%%%%%%%%%%%%%%%%%%%%%%%%%%%%
\section{\dots and the CMB\dots}

\begin{figure}[t]
\centering
\epsfig{figure=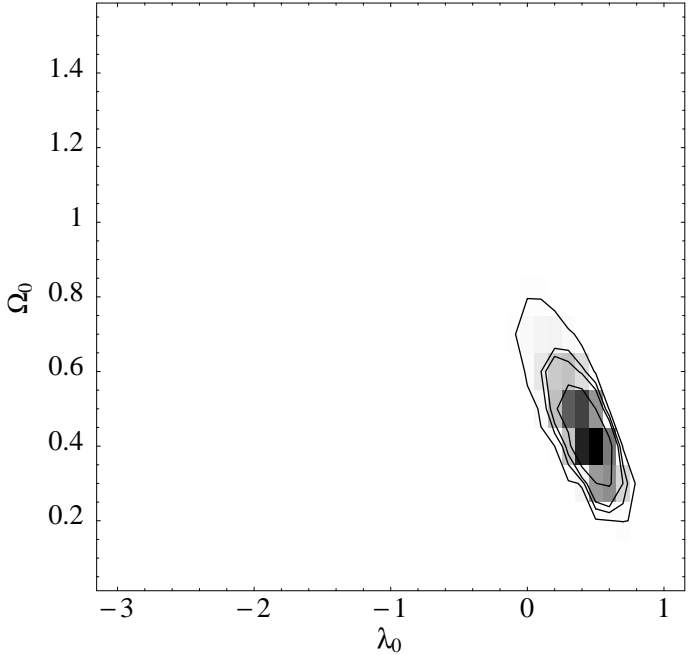, width=9cm}
\caption[]{
%\Large\bf\boldmath
The same as Fig.~\ref{fi:posterior} but 
combining \JVAS\, optical 
gravitational lens surveys from the literature and prior information on 
the value of $\Omega_{0}$, $H_{0}$ and the age of the universe with 
constraints derived from CMB anisotropies.  Since the CMB constraints 
are more or less orthogonal to the lensing statistics constraints, this 
reduces the allowed area of the $\lambda_{0}$-$\Omega_{0}$ parameter 
space significantly.  Note that the scale of this plot differs from the 
previous ones.  For technical reasons no models with 
$k=\sign(\lambda_{0}+\Omega_{0}-1)=+1$ were calculated; this slightly distorts 
the contours near the $k=0$ line, which otherwise would extend a bit 
more into the $k=+1$ region.}
\label{fi:cmb}
\end{figure}

It has long been realised,
e.g.~\cite{DEisensteinHT98b},
that the direction of degeneracy of constraints from cosmic microwave
background anisotropies is roughly orthogonal to that of most other
tests, including lensing statistics.  Thus, combining the constraints from 
CMB anisotropies with those
from other cosmological tests can give much tighter constraints than
either alone.  We have performed an analysis similar to that done in
\cite{CLineweaver98a}, though including an updated Tenerife data point and
calculating two-dimensional joint likelihood constraints as in the 
calculations done previously in this poster rather than employing Lineweaver's 
statistical method.  Adding the constraints on $\lambda_{0}$ and 
$\Omega_{0}$ so derived to the previous ones narrows down the region of
parameter space further, as is shown in Fig.~\ref{fi:cmb}.
%%%%%%%%%%%%%%%%%%%%%%%%%%%%%%%%%%%%%%%%%%%%%%%%%%%%%%%%%%%%%%%%%%%%%%%%
%%%%%%%%%%%%%%%%%%%%%%%%%%%%%%%%%%%%%%%%%%%%%%%%%%%%%%%%%%%%%%%%%%%%%%%%
%%%%%%%%%%%%%%%%%%%%%%%%%%%%%%%%%%%%%%%%%%%%%%%%%%%%%%%%%%%%%%%%%%%%%%%%

\begin{figure}[t]
\centering
\epsfig{figure=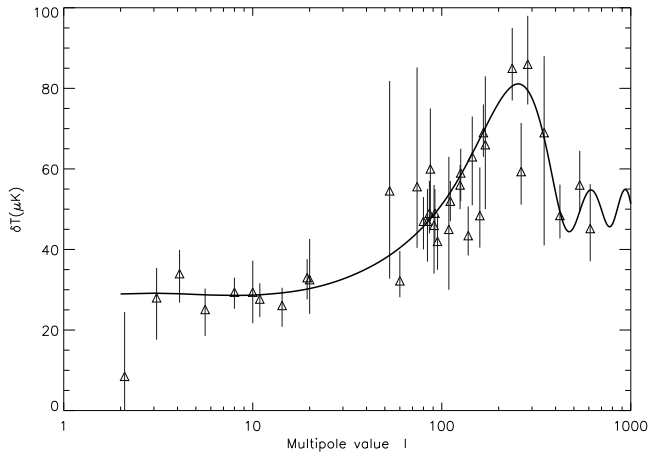, width=9cm}
\caption[]{
%\Large\bf\boldmath
Power spectrum shown with data points from the 
literature for 
our best-fit cosmological model, fitting to the CMB data alone (and 
keeping parameters 
other than $\lambda_{0}$ and $\Omega_{0}$ fixed at predetermined 
fiducial values).}
\label{fi:bestfit}
\end{figure}

The power spectrum for the best-fit model using the CMB data alone 
($\lambda_{0}=0.6$, $\Omega_{0}=0.3$, otherwise not shown here) along 
with various data points from the literature
(see the collection of Max Tegmark at 
\texttt{http://www.sns.ias.edu/\symbol{126}max/cmb/experiments.html})
is shown in Fig.~\ref{fi:bestfit}.
%%%%%%%%%%%%%%%%%%%%%%%%%%%%%%%%%%%%%%%%%%%%%%%%%%%%%%%%%%%%%%%%%%%%%%%%
%%%%%%%%%%%%%%%%%%%%%%%%%%%%%%%%%%%%%%%%%%%%%%%%%%%%%%%%%%%%%%%%%%%%%%%%
%%%%%%%%%%%%%%%%%%%%%%%%%%%%%%%%%%%%%%%%%%%%%%%%%%%%%%%%%%%%%%%%%%%%%%%%

(Note added for the proceedings.  The likelihood based on CMB
observations which (combined with other tests) is shown in
Fig.~\ref{fi:cmb} is, due to a numerical error, qualitatively but not
quantitatively correct.  We have since performed the correct
computation, which will be presented elsewhere.  However, the difference
creates a smaller error than that caused by many other approximations
made use of in these calculations, so we present the figure as shown in 
the original poster, in keeping with the concept of providing a record 
of the conference as opposed to an updated paper on the same subject.)

%%%%%%%%%%%%%%%%%%%%%%%%%%%%%%%%%%%%%%%%%%%%%%%%%%%%%%%%%%%%%%%%%%%%%%%%
%%%%%%%%%%%%%%%%%%%%%%%%%%%%%%%%%%%%%%%%%%%%%%%%%%%%%%%%%%%%%%%%%%%%%%%%
%%%%%%%%%%%%%%%%%%%%%%%%%%%%%%%%%%%%%%%%%%%%%%%%%%%%%%%%%%%%%%%%%%%%%%%%
\section{\dots and how this compares to other 
cosmological tests}

Fig.~\ref{fi:cmb} combines constraints based on optical and radio 
gravitational lensing statistics, `direct' measurements of $H_{0}$, 
$\Omega_{0}$ and the age of the universe and constraints derived from 
CMB anisotropies.  This restricts $\lambda_{0}$ to a narrow range.
If one believes that $\Omega_{0}\approx 0.3$, then it follows that
$\lambda_{0}\approx 0.5$.  This should be compared to the result
of Perlmutter et al.~\cite{SPerlmutteretal99a}:
$0.8\Omega_{0} - 0.6 \lambda_{0} \approx -0.2$: inserting $0.3$ for 
$\Omega_{0}$ one obtains $\lambda_{0}\approx0.43$.  Taking the errors 
into consideration (which are not large enough in either case to allow, 
for example, $\lambda_{0} = 0$) one obtains perfectly consistent 
measurements of $\lambda_{0}$ from completely independent methods.  

Taken together, present measurements of cosmological parameters
\emph{definitely} rule out the Einstein-de~Sitter universe
($\lambda_{0}=0$, $\Omega_{0}=1$), \emph{very probably} rule out a
universe without a cosmological constant ($\lambda_{0}=0$) and
\emph{tentatively} rule out a flat ($\lambda_{0}$ + $\Omega_{0}$ = 1)
universe as well.  A universe with $\lambda_{0}\approx 0.4$ and
$\Omega_{0}\approx 0.3$ seems to be consistent with all observational
data, including measurements of the Hubble constant and age of the
universe.

%%%%%%%%%%%%%%%%%%%%%%%%%%%%%%%%%%%%%%%%%%%%%%%%%%%%%%%%%%%%%%%%%%%%%%%%
%%%%%%%%%%%%%%%%%%%%%%%%%%%%%%%%%%%%%%%%%%%%%%%%%%%%%%%%%%%%%%%%%%%%%%%%
%%%%%%%%%%%%%%%%%%%%%%%%%%%%%%%%%%%%%%%%%%%%%%%%%%%%%%%%%%%%%%%%%%%%%%%%
\section{The future}

\CLASS\ is similar to \JVAS\ but contains about 4 times as many sources.  
The
definition of both is flat-spectrum between L-band and C-band,
i.e.~$\alpha>-0.5$ where $s_{f}\sim f^{\alpha}$, the essential 
difference
being the lower flux density limit of 200\,mJy for \JVAS\ and 30\,mJy for 
\CLASS.
However, since \CLASS\ is defined based on the newer GB6 and NVSS catalogues
\cite{PGregorySDJ96a,JCondonCGYPTB98a}
than \JVAS, there will be some essentially random differences due to
differing quality of observations and variability of the sources.
All the \JVAS\ lenses mentioned in Table~\ref{ta:lenses} are in
the new \CLASS\ sample, which, having no upper flux density limit, 
subsumes \JVAS.
The previous samples \CLASS-I and \CLASS-II will be similarly subsumed
in the same sense as \JVAS, though the differences here will be slightly
larger since bands other than L and C were used in the preliminary
definition of these samples.

The initial phase of observations is complete; currently lens candidates 
are being followed up.  At present, we have confirmed as gravitational 
lenses the systems listed in Table~\ref{ta:class} (which for 
completeness also includes the two \JVAS\ lens systems not used in the 
statistical analysis presented here).
%%%%%%%%%%%%%%%%%%%%%%%%%%%%%%%%%%%%%%%%%%%%%%%%%%%%%%%%%%%%%%%%%%%%%%%%
%%%%%%%%%%%%%%%%%%%%%%%%%%%%%%%%%%%%%%%%%%%%%%%%%%%%%%%%%%%%%%%%%%%%%%%%
\begin{table}[h]
\begin{center}
\begin{tabular*}{\textwidth}{@{\extracolsep{\fill}}llllll}
\hline
\hline
%Survey &
Name &
\# images &
$\Delta\theta''$ &
$z_{\mathrm{l}}$ &
$z_{\mathrm{s}}$ &
lens galaxy \\
\hline
%%_________________________________________________________________
%JVAS          &                % SURVEY
%B0218+357     &                % NAME
%2 + ring      &                % # IMAGES
%0.334         &                % IMAGE SEPARATION
%0.6847        &                % LENS REDSHIFT
%0.96          &                % SOURCE REDSHIFT
%spiral        \\               % LENS GALAXY TYPE
%%_________________________________________________________________
%JVAS          &                % SURVEY
%MG0414+054    &                % NAME
%4             &                % # IMAGES
%2.09          &                % IMAGE SEPARATION
%0.9584        &                % LENS REDSHIFT
%2.639         &                % SOURCE REDSHIFT
%elliptical    \\               % LENS GALAXY TYPE
%_________________________________________________________________
%CLASS         &                % SURVEY
B0712+472     &                % NAME
4             &                % # IMAGES
1.27          &                % IMAGE SEPARATION
0.406         &                % LENS REDSHIFT
1.34          &                % SOURCE REDSHIFT
spiral        \\               % LENS GALAXY TYPE
%%_________________________________________________________________
%JVAS          &                % SURVEY
%B1030+074     &                % NAME
%2             &                % # IMAGES
%1.56          &                % IMAGE SEPARATION
%0.599         &                % LENS REDSHIFT
%1.535         &                % SOURCE REDSHIFT
%spiral        \\               % LENS GALAXY TYPE
%_________________________________________________________________
%CLASS         &                % SURVEY
B1127+385     &                % NAME
2             &                % # IMAGES
0.70          &                % IMAGE SEPARATION
\textbf{?}    &                % LENS REDSHIFT
\textbf{?}    &                % SOURCE REDSHIFT
\textbf{?}    \\               % LENS GALAXY TYPE
%_________________________________________________________________
%JVAS          &                % SURVEY
%B1422+231     &                % NAME
%4             &                % # IMAGES
%1.28          &                % IMAGE SEPARATION
%0.337         &                % LENS REDSHIFT
%3.62          &                % SOURCE REDSHIFT
%\textbf{?}    \\               % LENS GALAXY TYPE
%_________________________________________________________________
%CLASS         &                % SURVEY
B1600+434     &                % NAME
2             &                % # IMAGES
1.39          &                % IMAGE SEPARATION
0.414         &                % LENS REDSHIFT
1.589         &                % SOURCE REDSHIFT
spiral        \\               % LENS GALAXY TYPE
%_________________________________________________________________
%CLASS         &                % SURVEY
B1608+656     &                % NAME
4             &                % # IMAGES
2.08          &                % IMAGE SEPARATION
0.63          &                % LENS REDSHIFT
1.39          &                % SOURCE REDSHIFT
spiral        \\               % LENS GALAXY TYPE
%_________________________________________________________________
%CLASS         &                % SURVEY
B1933+507     &                % NAME
$4+4+2$       &                % # IMAGES
1.17          &                % IMAGE SEPARATION
0.755         &                % LENS REDSHIFT
\textbf{?}    &                % SOURCE REDSHIFT
\textbf{?}    \\               % LENS GALAXY TYPE
%_________________________________________________________________
%JVAS          &                % SURVEY
B1938+666     &                % NAME
$4+2$         &                % # IMAGES
0.93          &                % IMAGE SEPARATION
\textbf{?}    &                % LENS REDSHIFT
\textbf{?}    &                % SOURCE REDSHIFT
\textbf{?}    \\               % LENS GALAXY TYPE
%_________________________________________________________________
%CLASS         &                % SURVEY
B2045+265     &                % NAME
4             &                % # IMAGES
1.86          &                % IMAGE SEPARATION
0.867         &                % LENS REDSHIFT
1.28          &                % SOURCE REDSHIFT
\textbf{?}    \\               % LENS GALAXY TYPE
%\hline
%_________________________________________________________________
%JVAS          &                % SURVEY
B2114+022     &                % NAME
2 or 4        &                % # IMAGES
2.57          &                % IMAGE SEPARATION
0.32 \& 0.59  &                % LENS REDSHIFT
\textbf{?}    &                % SOURCE REDSHIFT
\textbf{?}    \\               % LENS GALAXY TYPE
\hline
\end{tabular*}
\caption[]{
%\Large\bf\boldmath 
\CLASS\ gravitational lenses and the 
two \JVAS\ lens systems (1938+666 and 2114+022)
not listed in Table~\ref{ta:lenses}.}
\label{ta:class}
\end{center}
\end{table}
%%%%%%%%%%%%%%%%%%%%%%%%%%%%%%%%%%%%%%%%%%%%%%%%%%%%%%%%%%%%%%%%%%%%%%%%

We hope that the larger size of \CLASS\ will allow the constraints on
cosmological parameters from gravitational lensing statistics to
improve.  At present, the greatest uncertainty is the redshift-dependent 
luminosity function (or equivalently the flux-dependent redshift 
distribution) of the unlensed population (which of course, due to the 
amplification bias, extends to fainter flux-densities than the survey 
itself).  We are currently taking steps to decrease this uncertainty.

%%%%%%%%%%%%%%%%%%%%%%%%%%%%%%%%%%%%%%%%%%%%%%%%%%%%%%%%%%%%%%%%%%%%%%%%
%%%%%%%%%%%%%%%%%%%%%%%%%%%%%%%%%%%%%%%%%%%%%%%%%%%%%%%%%%%%%%%%%%%%%%%%
%%%%%%%%%%%%%%%%%%%%%%%%%%%%%%%%%%%%%%%%%%%%%%%%%%%%%%%%%%%%%%%%%%%%%%%%
\section*{Acknowledgements}

{\centering
\epsfig{figure=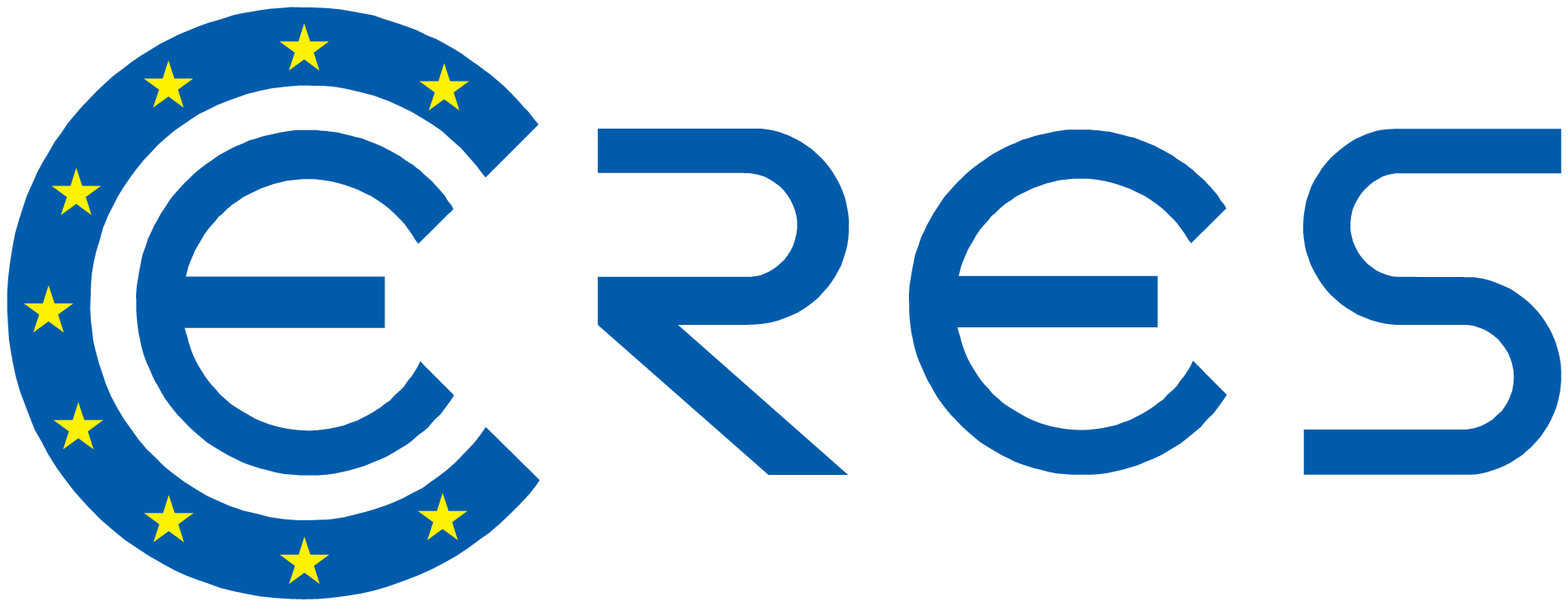, width=9cm}
}

\noindent
This research was supported in part by the European Commission, TMR
Programme, Research Network Contract ERBFMRXCT96-0034 `CERES'.  \CERES,
standing for the Consortium for European Research on Extragalactic 
Surveys,
is an EU TMR Network, coordinated by Ian Browne at Jodrell Bank, and 
involving the Nuffield Radio Astronomy Laboratories of the Universtity 
of Manchester at Jodrell Bank, the Institute of Astronomy of the 
University of Cambridge, 
The Kapteyn Astronomical Institute of the University of Groningen, the
Netherlands Foundation for Research in Astronomy at Dwingeloo, the 
University of Bologna and the University of Portugal.  One of the main 
goals of \CERES\ is the use of the \JVAS\ and \CLASS\ surveys for 
research in gravitational lensing.

We thank
our collaborators in the \JVAS, CJF and \CLASS\ surveys for useful
discussions and for providing data in advance of publication and many
colleagues
at Jodrell Bank for helpful comments and suggestions.  We also
thank John Meaburn and Anthony Holloway at the Department of Astronomy
in Manchester and the staff at Manchester Computing for providing us
with additional computational resources.  RQ is grateful to
the \CERES\ collaboration for making possible a visit to Jodrell Bank
where part of this work was done.

%%%%%%%%%%%%%%%%%%%%%%%%%%%%%%%%%%%%%%%%%%%%%%%%%%%%%%%%%%%%%%%%%%%%%%%%
%%%%%%%%%%%%%%%%%%%%%%%%%%%%%%%%%%%%%%%%%%%%%%%%%%%%%%%%%%%%%%%%%%%%%%%%
%%%%%%%%%%%%%%%%%%%%%%%%%%%%%%%%%%%%%%%%%%%%%%%%%%%%%%%%%%%%%%%%%%%%%%%%

\vspace{1ex}

\noindent
\CERES\ publications: 
\texttt{http://multivac.jb.man.ac.uk:8000/ceres/papers/papers.html}

\end{document}